# 2 TO 16 GHZ FUNDAMENTAL SYMMETRIC MODE ACOUSTIC RESONATORS IN PIEZOELECTRIC THIN-FILM LITHIUM NIOBATE

*Vakhtang Chulukhadze, Jack Kramer, Tzu-Hsuan Hsu, Omar Barrera, Ian Anderson, Sinwoo Cho,*
*Joshua Campbell, and Ruochen Lu*
The University of Texas at Austin, Austin, Texas, USA

## ABSTRACT

As 5G connectivity proliferates, signal processing applications at 6G centimeter bands have gained attention for urban wireless capacity expansion. At sub-5 GHz, acoustic resonators operating in the fundamental symmetric (S0) Lamb mode hold significant promise if frequency scaled to the 6G centimeter bands. Concurrently, the lateral wavelength ($\Lambda$) dependency and the traveling wave nature of S0 mode enable monolithic multi-frequency fabrication, transversal filters, correlators, and other compact signal processing components. In this work, we present thin-film lithium niobate (LN) S0 resonators scaled up to 16 GHz. Specifically, we study the characteristics of the S0 mode as $\Lambda$ is minimized and showcase a device at 14.9 GHz with a Bode Q maximum ($Q_{max}$) of 391, a $k^2$ of 6%, and a figure of merit (FoM) of 23.33 – surpassing the state-of-the-art (SoA) in its frequency range [Fig. 8].

## INTRODUCTION

As 5G connectivity proliferates, signal processing applications at 6G centimeter bands have gained attention for urban wireless capacity expansion [1]–[4]. At lower frequencies, the development of high FoM acoustic resonators has been an important facilitator of radio frequency (RF) technology as acoustic devices possess a smaller footprint and lower loss compared to their electro-magnetic (EM) counterparts [5]–[7]. Materials such as lithium niobate (LN), aluminum nitride (AlN), and scandium doped aluminum nitride have been integral to this success [8], [9]. These piezoelectric materials possess high electro-mechanical coupling ($k^2$) for a variety of acoustic modes, and their high crystalline quality ensures quality factors ($Q$) suitable for low-loss operation. Aluminum Nitride has proven to be highly successful due to its moderate $k^2$, high $Q$ and CMOS compatibility. Similarly, LN has seen significant success due to its extremely high $k^2$ and high $Q$. However, conventional LN, AlN, and ScAlN acoustic front-end signal processing platforms, currently dominated by Surface Acoustic Wave, and Film Bulk Acoustic Wave technologies, are either difficult to scale in frequency, or are associated with a complex fabrication process [10], [11]. Meanwhile, acoustic resonators operating in the fundamental symmetric (S0) Lamb mode have shown high quality $Q$, $k^2$, and figure-of-merit (FoM, $k^2 \cdot Q$) whilst avoiding fabrication difficulties associated with a bottom electrode – S0 mode is a promising candidate for frequency scaling LN and AlN towards 6G centimeter wave bands [12], [13]. Moreover, the lateral wavelength ($\Lambda$) dependency of S0 mode enables monolithic multi-frequency fabrication as well as high frequency precision – a key issue concerning FBAR technology. At the same time, S0 mode is a travelling wave [Fig. 1], hence, transversal filters, correlators, acoustic delay lines, and a variety of other compact signal processing components become viable through the development of this platform [14]. In this work, we present thin-film lithium niobate S0 resonators scaled up to 16 GHz. We study the characteristics of the S0 mode as $\Lambda$ is minimized, paving the way for future success.

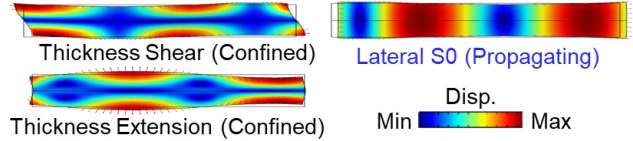

*Figure 1: An illustration depicting the traveling wave nature of the S0 mode in comparison with the confined thickness shear and thickness extensional modes in LN*

At the same time, we readily showcase multiple devices surpassing the state-of-the-art (SoA) in their frequency range, indicating towards the scalability of acoustics to 6G centimeter wave bands. More specifically, we showcase an acoustic resonator at 14.9 GHz with a Bode Q maximum ($Q_{max}$) of 391, a $k^2$ of 6%, and FoM of 23.33 – surpassing the state-of-the-art (SoA) in its frequency range [Fig. 8].

## MATERIAL SELECTION

RF Acoustics largely relies on the strong piezoelectricity exhibited by LN, AlN, and ScAlN. All three materials possess strong potential for low-loss and high frequency operation, however, the choice between them must still be made carefully. While it does not possess $k^2$ as large as its competitors, AlN is CMOS compatible, and its deposition at high quality has been well-explored [15], [16]. ScAlN improves upon AlN by enhancing available $k^2$ through scandium doping, however, its deposition at high quality and optimum doping remains a subject of active research [17]. LN, unlike AlN and ScAlN is not CMOS compatible, however, it has an incredibly high $k^2$ for a variety of acoustic modes while also possessing a high $f \cdot Q$ product, a metric commonly used for assessing piezoelectric material quality [18]. Importantly, the process of fabricating a piezoelectric stack using LN through a wafer transfer process has been well researched [19]. As a result, the potential frequency scalability of LN has been well-recognized, and LN is considered a most promising piezoelectric platform for scaling acoustics to beyond 5 GHz [3]. Hence, we have chosen LN as our material of choice whilst scaling acoustics to 6G centimeter wave bands.

LN is a highly anisotropic single-crystal material. Accordingly, its material properties change in different orientations and allow varying levels of $k^2$ for the desired acoustic mode. Hence, the appropriate crystal cut must be identified to frequency scale the S0 mode while avoiding detrimental effects like spurious modes and low $k^2$. The appropriate piezoelectric electro-mechanical coupling matrix ($e$) coefficient for the longitudinal motion characteristic of the S0 mode is $e_{11}$. Considering the commercial availability of different LN orientations containing a high $e_{11}$, our choices are restricted to X-cut, Y-cut, and 128Y LN. While X-cut shows superior $k^2$ for the S0 mode, it also contains large amounts of parasitic coupling – moderate $e_{13}$ and $e_{16}$, leading to spurious modes. Meanwhile, 128Y LN has slightly lower $e_{11}$ compared to X-cut LN, but it is still almost double the magnitude of $e_{11}$ available in Y-cut LN. Moreover 128Y LN contains minimal parasitic coupling at the in-plane orientation at which $e_{11}$ is maximized. Hence, 128Y LN is a suitable choice for studying the characteristics of the S0 mode.

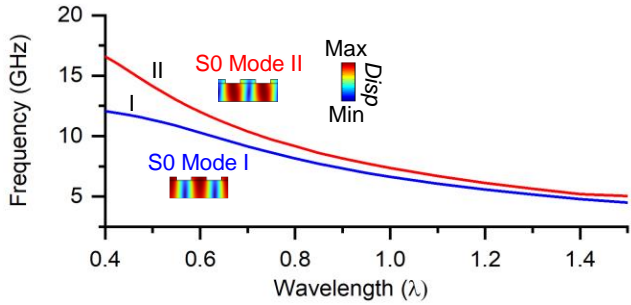

*Figure 2: Dispersion analysis of a S0 mode resonator in LN, showcasing mode-splitting behavior*

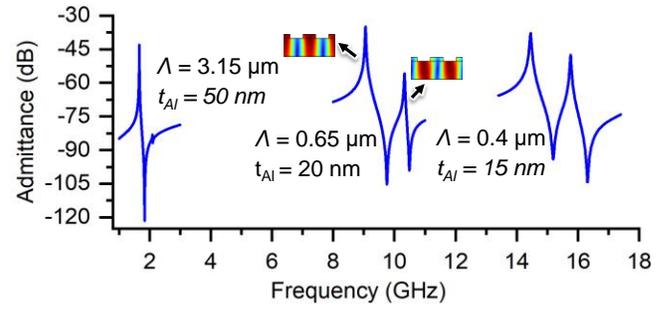

*Figure 4: Simulated admittance graphs for S0 mode devices given the frequency scaling plan whilst accounting for the shadowing effect*

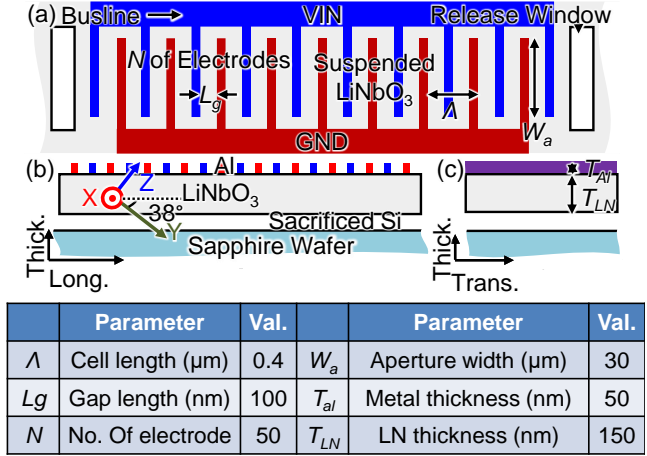

*Figure 3: (a) top-down view of device design, (b) specific design parameters for the highest-frequency device*

Accordingly, a 110 nm thick 128Y LN on amorphous silicon (aSi) on sapphire ($Al_2O_3$) was provided by NGK insulators ltd.

## DESIGN AND FABRICATION

The resonant frequency of the S0 mode is defined by the lateral wavelength corresponding to the interdigitated electrodes (IDT). It follows that the definition of nanometer-scale IDT patterns with electron beam lithography (EBL) can be considered as the most challenging fabrication step. With this limitation in mind, device $\Lambda$ was varied from 0.4 to 3.15 μm to scale S0 mode devices to 6G centimeter wave bands, leading to a minimum feature size of 100 nm while the maximum feature size was limited to 787.5 nm. Next, the contrast between conventional high-frequency acoustic modes and the S0 mode was considered [Fig. 1]. Evidently, the S0 mode is a travelling wave, hence, the structural energy confinement will significantly influence the performance. Though thin and long anchors would be the natural choice to enhance this feature, they would likely cause the thin LN membrane to collapse, detrimental for structural energy confinement of the system. Accordingly, to ensure the mechanical robustness, a fully anchored design was chosen and implemented. Finally, an optimal IDT thickness was chosen while keeping certain considerations in mind. Specifically, the effects of miniature feature size on the metal deposition process, as well as the strong dispersion of LN. Firstly, from prior testing, a shadowing effect was expected at low $\Lambda$ – the metal deposition rate would decay for increasingly narrower patterns. Moreover, mode-splitting behavior was observed whilst performing dispersion analysis to study the effects of the metal/piezo thickness ratio for the S0 mode [Fig. 2]. Consequently, S0 mode would exist at two distinct frequencies – concentrated in the metal/piezo stack (S0 I) and in the piezo layer alone (S0 II). Hence, to minimize mode-splitting behavior, as well as account for the shadowing effect, a 50 nm electrode thickness was chosen and was assumed to decay with $\Lambda$. The top-view of the resultant device design, along with specific parameters can be seen on Fig. 3. Finally, the proposed frequency scaling plan was validated using COMSOL finite element analysis depicted on Fig. 4.

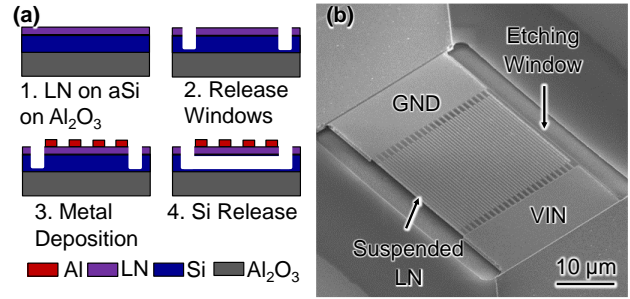

*Figure 5: (a) The proposed fabrication flow (b) Scanning electron microscope images of a sample fabricated device*

Release windows were first formed on a 128Y LN/aSi/Sapphire stack via a photo-lithography process, and an AJA ion milling tool. AZ5214 photoresist was patterned using an SUSS MicroTech MA-8 UV Exposure tool. The etch was performed with a 20° incline from normal incidence, at a low ion beam power of 400 V and a moderate argon gas flow of 6 sccm. The process was followed by a low-power smoothing step at 75° incline from normal to ensure sub-nm surface roughness and avoid the material re-deposition effect. Moreover, crystal bond was used as an adhesive layer to the actively cooled carrier wafer to mitigate high temperature effects which accumulate due to the low etch rate of LN. Having formed release windows, probing pads and IDTs were patterned in separate layers using EBL tools, and metal was deposited in two steps – 50 nm of Aluminum (Al) for the IDT and busline region, and 300 nm of Al for the probing pads. Finally, the aSi sacrificial layer was etched away using a Xenon Difluoride ($XeF_2$) etching tool, and the resonators were successfully released. A detailed fabrication flow can be seen on Fig. 5 (a), and a scanning electron microscope image of a fabricated device on Fig. 5 (b).

## RESULTS AND DISCUSSION

The measured response of sample devices is plotted in Fig 6, showcasing strong performance from 2 to 16 GHz. A specific highlight at high frequency includes a device with a $Q_{max}$ of 391, $k^2$ of 6%, and FoM of 23.33 at 14.9 GHz. The measurements showcase clear trends, the analysis of which will enable us to improve upon

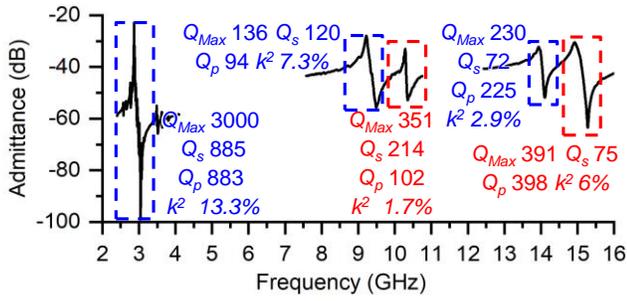

*Figure 6: The measured electrical response of sample fabricated devices*

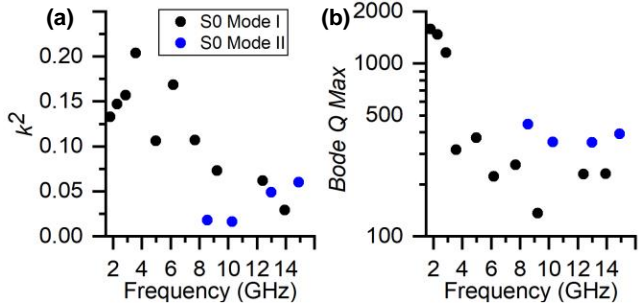

*Figure 7: (a) $k^2$ for S0 I and S0 II modes as calculated from the series and parallel resonance peaks (b) Bode $Q_{max}$ for S0 I and S0 II modes from 2-17 GHz on a logarithmic scale*

the results seen here, and effectively scale S0 mode devices to 6G centimeter wave bands.

A closer look at the performance of S0 I and S0 II modes against frequency is particularly interesting. Firstly, the concentration of electrical energy splits between the two modes as lambda is reduced [Fig. 7 (a)]. S0 I mode has considerably higher $k^2$ compared to S0 II at high $\Lambda$, which also corresponds to higher metal thickness – a consequence of the shadowing effect. Meanwhile, at high frequencies, the $k^2$ of the S0 I mode is lower at the cost of higher $k^2$ for the S0 II mode, leading to equalized electrical energy distribution between the two modes. At lower frequencies, the $k^2$ associated with the S0 II mode is miniscule, and is omitted from further analysis. However, for the high-frequency devices, $Q_{max}$ of S0 I and II can be directly compared – S0 II consistently outperforms S0 I while possessing almost equivalent $k^2$ [Fig. 7 (b)]. Accordingly, to further improve upon the results presented here at high frequency, we must ensure that we suppress the mode-splitting behavior in favor of the S0 II mode.

On another note, the discrepancy between the resonators' parallel resonance quality factor ($Q_p$) and their quality factor at series resonance ($Q_s$) provides some insight into the consequences of the device design. At low frequencies, the resonators show $Q_p$ and $Q_s$ which are very close together in value. Hence, the 50 nm electrode thickness leads to strong mechanical performance, and it is not too lossy electrically. However, the gap between $Q_s$ and $Q_p$ increases significantly as lambda is reduced. $Q_s$ decreases dramatically, which further confirms the shadowing effect – the resistive loss increases with frequency due to a thinner metal layer. The contrast between $Q_{max}$, $Q_s$, and $Q_p$ provides a similar insight – $Q_{max}$ can be found half-way between series and parallel resonances of the device at lower frequencies in the absence of spurious modes, however, $Q_{max}$ always coincides with $Q_p$ at low $\Lambda$, indicating towards high series resistance ($R_s$), and accordingly, excessive electrical loading near the series resonance peak. Again, this points not only to increasing electrical loss at higher frequency, but it also confirms the shadowing effect. Hence, future design of high-frequency S0 resonators must either restrict $\Lambda$ to achieve better IDT thickness uniformity and balanced performance, or deposit additional metal in successive sessions until uniformity has been achieved between low and high $\Lambda$.

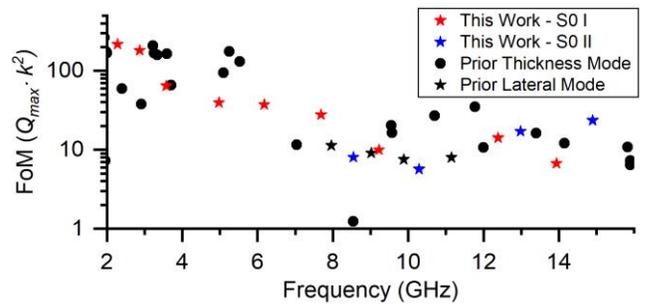

*Figure 8: Comparison of S0 mode devices to the SoA from 2 to 17 GHz, showcasing the potential of the S0 mode at high frequency* [9], [20]–[25]

## CONCLUSION

In this work, we have scaled the S0 mode to 6G Centimeter Wave bands, up to 16 GHz. We studied characteristics of the S0 mode as its operating frequency was increased, observed the mode-splitting behavior, along with the shadowing effect. Through this study, many of the inadvertent consequences of frequency scaling have been explored, and a clear path forward can be outlined to further improve upon the results of this work : high-frequency S0 mode resonators must suppress the S0 I mode for low-loss operation. A potential viable path to do so could be to fabricate devices using recessed electrodes with a high metal/piezo thickness ratio. Even at this stage of development, we report several acoustic resonators which out-perform the SoA in their frequency range, depicted on Figure 8. More specifically, this work surpasses concurrent lateral $\Lambda$ dependent devices while closely competing with, or surpassing thickness defined modes. This is an important point – S0 mode devices out-perform SAW beyond 6 GHz while they compete closely, or even surpass FBAR and XBAR technology reported from 6-16 GHz. Hence, with further development and mode-splitting suppression, S0 mode resonators could enable efficient urban capacity expansion for wireless communication in 6G centimeter wave bands.

## ACKNOWLEDGEMENTS

The author would like to thank the DARPA OPTIM and COFFEE projects for funding this work.